\documentclass[aps,
prd,
twocolumn,
nofootinbib
]{revtex4-1}

\usepackage
{graphicx}
\usepackage{amssymb}
\usepackage{amsmath}
\usepackage{braket}
\usepackage{listings}
\usepackage{cases}
\usepackage{comment}
\usepackage[colorlinks=true,
urlcolor=blue,
anchorcolor=blue,
citecolor=blue,
filecolor=blue,
linkcolor=blue,
menucolor=blue,
linktocpage=true,
pdfproducer=medialab,
pdfa=true
]{hyperref}

\def\slashchar#1{\setbox0=\hbox{$#1$} 
\dimen0=\wd0 
\setbox1=\hbox{/} \dimen1=\wd1 
\ifdim\dimen0>\dimen1 
\rlap{\hbox to \dimen0{\hfil/\hfil}} 
#1 
\else 
\rlap{\hbox to \dimen1{\hfil$#1$\hfil}} 
/ 
\fi}

\newcommand{\Mev}{\mathrm{MeV}}
\newcommand{\Gev}{\mathrm{GeV}}

\newcommand{\dif}[2]{\frac{\mathrm{d} #1}{\mathrm{d} #2}}

\newcommand{\dd}{\mathrm{d}}
\newcommand{\ee}{\mathrm{e}}

\begin{document}

\title{Revisiting constraints on small scale perturbations from big-bang nucleosynthesis}
\date{\today}
\author{Keisuke Inomata}
\email{inomata@icrr.u-tokyo.ac.jp}
\author{Masahiro Kawasaki}
\email{kawasaki@icrr.u-tokyo.ac.jp}
\author{Yuichiro Tada}
\email{yuichiro.tada@ipmu.jp}
\affiliation{Institute for Cosmic Ray Research, The University of Tokyo, Kashiwa, Chiba 277-8582, Japan}
\affiliation{Kavli Institute for the Physics and Mathematics of the Universe (WPI), UTIAS, The University of Tokyo, Kashiwa, Chiba 277-8583, Japan}

\preprint{IPMU 16-0068}

\begin{abstract}
We revisit the constraints on the small scale density perturbations ($10^4\,\mathrm{Mpc}^{-1}\lesssim k \lesssim10^5\,\mathrm{Mpc}^{-1}$) 
from the modification of the freeze-out value of the neutron-proton ratio at big-bang nucleosynthesis era. 
Around the freeze-out temperature $T\sim 0.5\,\mathrm{MeV}$, the universe can be divided into several local patches which have different temperatures since
any perturbation which enters the horizon after the neutrino decoupling has not diffused yet.
Taking account of this situation, we calculate the freeze-out value in detail.
We find that the small scale perturbations decrease the n-p ratio in contrast to previous works. 
With use of the latest observed $^4$He abundance, we obtain the constraint on the power spectrum of the curvature perturbations as
$\Delta^2_\mathcal{R}\lesssim 0.018$ on $10^4\,\mathrm{Mpc}^{-1}\lesssim  k \lesssim 10^5\,\mathrm{Mpc}^{-1}$.
\end{abstract}

\maketitle

\section{Introduction}

In recent decades, modern cosmology has got into a phase of highly precise physics. 
Especially progress of observations of the cosmic microwave background (CMB) is quite remarkable, and recent results of Planck collaboration are beautifully consistent with the standard $\Lambda$CDM scenario~\cite{Ade:2015xua}. 
On the other hand, success of the big-bang nucleosynthesis (BBN) has supported the standard big-bang model of cosmology for more than 50 years and BBN also has been used as an important probe to physical states of the early universe.
Together with  measurements of abundances of light elements (e.g. D and  $^4$He)
BBN can give stringent constraints on various cosmological scenarios. 
The latest observed value of the primordial 
${}^4$He and D abundances, $Y_p^\mathrm{obs}=0.2449\pm0.0080~(2\sigma)$~\cite{Aver:2015iza}  and $(\text{D/H})^\text{obs}_p= (2.53 \pm 0.04) \times 10^{-5}$~\cite{Cooke:2013cba}, are in good agreement with the standard BBN prediction with the best fit parameters of Planck, $Y_p^\mathrm{CMB}=0.24668\pm0.00013$ and $(\text{D/H})^\text{CMB}_p = 2.606^{+0.051}_{-0.054} \times 10^{-5}$~\cite{Ade:2015xua}.

As an initial condition of big-bang cosmology, inflation, i.e. the accelerated expansion of the early universe, is one of the most plausible scenario. 
Inflation can not only solve several problems of big-bang theory such as horizon, flatness, and monopole problems, but also create primordial inhomogeneities from quantum fluctuations as seeds of cosmic structures. 
That is, the information of the dynamics of the early universe will be imprinted on cosmological inhomogeneities like the large scale structure (LSS) and the temperature and polarization anisotropies of CMB.

From observations of CMB and LSS, the primordial curvature perturbations are found to be almost scale invariant and the amplitude of their power spectrum $\Delta_\mathcal{R}^2(k)$ is as small as $\mathcal{O}(10^{-9})$ at least on large scales such that $k\lesssim \mathcal{O}(1\,\mathrm{Mpc}^{-1})$~\cite{Ade:2015xua,Ade:2015lrj}.
However, they are still not understood well on smaller scales $k\gtrsim \mathcal{O}(1\,\mathrm{Mpc}^{-1})$ because of the Silk damping \cite{Silk:1967kq}. 
Theoretically, there are many inflationary models which predict the large enhancement of the small scale perturbations, e.g. \emph{simple hybrid inflation}~\cite{GarciaBellido:1996qt,Randall:1995dj} (see also \cite{Fujita:2014tja,Clesse:2015wea,Kawasaki:2015ppx} for concrete power spectra), \emph{hybrid-new double inflation}~\cite{Kawasaki:1997ju,Kawaguchi:2007fz,Frampton:2010sw}, \emph{single-field double inflation}~\cite{Yokoyama:1998pt,Bugaev:2008bi}, \emph{running mass model}~\cite{Kohri:2007qn,Drees:2011hb}, and \emph{curvaton scenario}~\cite{Kawasaki:2012wr,Kohri:2012yw}.
Therefore it is important to investigate the small scale perturbations to understand the early universe.

In this context, several methods to probe the small scale perturbations are studied. 
Primordial Black holes (PBHs)~\cite{Hawking:1971ei,Carr:1974nx,Carr:1975qj}, which are suggested to be produced by the gravitational collapse of overdense Hubble patches in the radiation dominated era, are one of the representative probes.
PBHs have been not detected yet and the non-detection of PBHs can constrain the primordial curvature perturbations for a very wide range of the wavelength as $10^{-2}\,\mathrm{Mpc}^{-1}\lesssim k\lesssim 10^{19}\,\mathrm{Mpc}^{-1}$ but weakly $\Delta_\mathcal{R}^2\lesssim \mathcal{O}(10^{-2})$~\cite{Josan:2009qn,Carr:2009jm}.
As another collapsed object, ultracompact minihalos (UCMHs)~\cite{Ricotti:2009bs} also receive notable attention. 
Non-detection of gamma rays from dark matter (DM) annihilations in UCMHs puts much stronger constraints $\Delta_\mathcal{R}^2\lesssim \mathcal{O}(10^{-7})$ on scales of $10\,\mathrm{Mpc}^{-1}\lesssim  k\lesssim 10^7\,\mathrm{Mpc}^{-1}$.
However, this constraint only applies to WIMP-DM (Weakly Interacting Massive Particles) which is relatively massive $\sim\mathcal{O}(1\,\mathrm{TeV})$ and has a weak charge so that the cross section is around $\braket{\sigma v}\sim10^{-26}\,\mathrm{cm}^3\mathrm{s}^{-1}$~\cite{Bringmann:2011ut}.\footnote{%
Recently it has been suggested that the UCMH abundance can be constrained by pulser timing irrespectively of
the DM properties~\cite{Clark:2015sha,Aslanyan:2015hmi}.}
The constraints from CMB spectral distortions are around $\Delta_\mathcal{R}^2\lesssim \mathcal{O}(10^{-5})$ on relatively large scales as $1\,\mathrm{Mpc}^{-1}\lesssim  k \lesssim 10^4\,\mathrm{Mpc}^{-1}$~\cite{Fixsen:1996nj,Hu:1994bz,Chluba:2012gq,Chluba:2012we,Chluba:2013dna} and they will be much improved by future CMB observations like PIXIE~\cite{Kogut:2011xw} or PRISM~\cite{Andre:2013afa}.

Recently, the idea of \emph{acoustic reheating} has been proposed as a new method to probe the perturbations on $10^4\,\mathrm{Mpc}^{-1}\lesssim  k\lesssim 10^5\,\mathrm{Mpc}^{-1}$, which is slightly smaller than the CMB distortion scale, by Jeong et al.~\cite{Jeong:2014gna} and Nakama et al.~\cite{Nakama:2014vla}.
This scale corresponds with the horizon scales around the BBN phase and the constraints come from the possibility that large density perturbations can affect the abundances of light elements.
Nakama et al. estimated the constraint as $\Delta_\mathcal{R}^2\lesssim 0.06$ from the modulation of the baryon-photon ratio $\eta=n_b/n_\gamma$ where $n_b$ and $n_\gamma$ represent number densities of 
baryons and photons respectively. On the other hand, Jeong et al. put the constraint of $\Delta_\mathcal{R}^2\lesssim 0.007$ from the increase of the ${}^4$He abundance $Y_p$ with perturbations. 

In this paper, we revisit these constraints, focusing especially on the modification of the freeze-out value of the neutron-proton ratio, which determines the primordial ${}^4$He abundance dominantly. 
Taking account of the fact that 
perturbations of electrons and neutrinos which enter the horizon after the neutrino decoupling evolve still together with baryon-photon fluid during that era,
it is found that the perturbations decrease the ${}^4$He abundance oppositely to Jeong et al.~\cite{Jeong:2014gna}
and the resultant constraint on the primordial curvature perturbations is $\Delta_\mathcal{R}^2\lesssim 0.018$.
These modifications are caused by second order effects of perturbations. We evaluate them in the iterative approximation in this paper, that is, we use the linear order solutions for the perturbations
and approximate the second order quantities by cross terms of linear perturbations without solving exact second order equations of motion (E.o.M).

This paper is organized as follows. 
In section~\ref{small scale perturbations}, we clarify the setup and review the dynamics of the cosmological perturbations around the horizon scale.
We show that indeed the decaying modes cannot be neglected yet and the relation between the temperature perturbations and the primordial curvature perturbations is modified by several factors. 
Then we calculate the n-p ratio with perturbations and find the constraint on the primordial curvature perturbations in section~\ref{n-p ratio}. 
We devoted section~\ref{conclusions} to the conclusions.

\section{Small scale perturbations in BBN epoch}
\label{small scale perturbations}

In this section we discuss the evolution of the small scale perturbations in BBN epoch. 
In particular, we focus on the epoch when the n-p ratio freezes out ($0.1\,\mathrm{MeV}\lesssim  T\lesssim 2\,\mathrm{MeV}$).

\subsection{Behavior of electrons and neutrinos}

As we will describe in detail in the next section,
the freeze-out value of the n-p ratio is mainly determined by the distributions of electrons and neutrinos.
During this epoch, electrons remain tightly coupled with the baryon-photon plasma and their perturbations have not been dissipated by the Silk damping. 
On the other hand, the situation of neutrinos is different. 
At high temperature ($T>1.5\,\mathrm{MeV}$), neutrinos are tightly coupled with baryon-photon plasma as well as electrons. 
However, when the radiation temperature drops to around $1.5\,\mathrm{MeV}$, neutrinos decouple from the baryon-photon plasma~\cite{Mukhanov:2005} and their free-streaming length reaches the horizon scale. 
Moreover, since neutrinos start to free-stream gradually before the decoupling, they diffuse almost all of the subhorizon perturbations of the baryon-photon plasma, too~\cite{Jedamzik:1996wp}. 
Namely just before the decoupling all density perturbations smaller than $k\sim10^5\,\mathrm{Mpc}^{-1}$ which is the horizon scale at the decoupling are erased and rethermalized.

Meanwhile superhorizon perutrbations have not been erased yet. Let us consider the perturbations which reenter the horizon slightly after the neutrino decoupling and before the freeze-out of the n-p ratio, 
that is, whose scales are given by $10^4\,\mathrm{Mpc}^{-1}\lesssim  k \lesssim 10^5\,\mathrm{Mpc}^{-1}$. 
After they reenter the horizon, the perturbations of electrons and neutrinos start to oscillate together with the baryon-photon plasma at first. 
After the first half oscillation, only the perturbations of neutrinos are smoothed off due to their free-streaming~\cite{Lesgourgues:2006nd}. 
However, the freeze out of the n-p ratio occurs before the first half oscillation as we will show in the next subsection 
and the perturbations are not yet smoothed off at that time. Therefore we do not have to consider the difference between electrons and neutrinos when we calculate the n-p ratio with the perturbations on $10^4\,\mathrm{Mpc}^{-1}\lesssim  k \lesssim 10^5\,\mathrm{Mpc}^{-1}$.

\subsection{Perturbations outside or inside horizon}

Now, let us consider the behavior of the perturbations for $10^4\,\mathrm{Mpc}^{-1}\lesssim  k\lesssim 10^5\,\mathrm{Mpc}^{-1}$. 
Hereafter we basically follow the notation of~\cite{Dodelson:2003}.
We use the conformal Newtonian gauge where the metric is given by, 
\begin{equation}
   \dd s^2= -a^2(1+2\Psi)\dd\eta^2+a^2(1+2\Phi)(\dd x^2+\dd y^2+\dd z^2)~.
\end{equation}
The radiation (photons and neutrinos) perturbations $\Theta$ can be defined by its distribution function $f$ as,
\begin{align}
	f(\vec{x},p,\hat{p},t)=\left[\exp\left(\frac{p}{T(t)[1+\Theta(\vec{x},\hat{p},t)]}\right)-1\right]^{-1}.
\end{align}
Here $p$ and $\hat{p}$ denote the amplitude and direction of the momentum of photons respectively.
Its Fourier mode $\Theta(\vec{k},p,\hat{p},t)=\int\dd^3x\,\ee^{-i\vec{x}\cdot\vec{k}}\Theta(\vec{x},p,\hat{p},t)$ is often used in Legendre expanded forms as,
\begin{align}
	\Theta_l(k,t)=\frac{1}{(-i)^l} \int^1_{-1} \frac{\dd\mu}{2} \emph{P}_l(\mu)\Theta(\vec{k},p,\hat{p},t), \quad \mu=\frac{\vec{k}\cdot\hat{p}}{k},
\end{align}
where $P_l$ is the Legendre polynomial of order $l$. 
Here note that after $\mu$-integration $\Theta_l$ does not dependent on $\vec{k}$-direction because of the homogeneity and isotropy of the universe.
Hereafter we omit the arguments of $\Theta$ for simplicity where it does not lead to confusion.
With use of the properties of the Legendre polynomial, we can obtain,
\begin{align}
	\int^1_{-1}\frac{\dd\mu}{2} \Theta^2&=\sum_{l=0}(2l+1)\Theta^2_l \nonumber \\
	&\simeq \Theta_0^2 + 3\Theta_1^2, \quad \text{(tight-coupling limit)}.
\end{align}
In the second line, we used the fact that in the tight-coupling limit\footnote{
When the freeze out is occurred, neutrinos are decoupled but oscillate together with the baryon-photon plasma. Therefore we can apply the tight-coupling limit to neutrinos as well as photons at that time.
}
 we can neglect $\Theta_l$ ($l>2$)~\cite{Dodelson:2003}. 
In the followings, we refer to $\Theta_0^2+3\Theta_1^2$ as $\bar{\Theta}^2$.

Assuming photons and neutrinos evolve conjointly until the freeze out of the n-p ratio, we can obtain the evolution of the gravitational potential ($\Phi$) and 
the radiation (photons and neutrinos) perturbations ($\Theta_0$ and $\Theta_1$) as~\cite{Dodelson:2003},
\begin{align}
	\label{g_potential}
	\Phi&=-3\Phi_p\left(\left(\frac{\sqrt{3}}{k\eta}\right)^2\cos\frac{k\eta}{\sqrt{3}}-\left(\frac{\sqrt{3}}{k\eta}\right)^3\sin\frac{k\eta}{\sqrt{3}}\right), \\
	\label{theta0}
	\Theta_0&=-\frac{3}{2}\Phi_p\left(\cos\frac{k\eta}{\sqrt{3}}-\frac{\sqrt{3}}{k\eta}\sin\frac{k\eta}{\sqrt{3}}\right)-\frac{3}{k\eta}\Theta_1, \\
	\label{theta1}
	\Theta_1&\!=\!-\frac{\sqrt{3}}{2}\Phi_p\!\left(\!\sin\frac{k\eta}{\sqrt{3}}\!+\!2\frac{\sqrt{3}}{k\eta}\cos\frac{k\eta}{\sqrt{3}}\!-\!2\!\left(\!\frac{\sqrt{3}}{k\eta}\!\right)^2\!\sin\frac{k\eta}{\sqrt{3}}\!\right)\!,
\end{align}
where $\Phi_p$ is an initial condition of the gravitational potential on the superhorizon scale and connected with the primordial curvature perturbation $\zeta$ by $\Phi_p=\frac{2}{3}\zeta$.\footnote{
This initial condition may have to be replaced by $\Phi_p=\frac{2}{3}\frac{1+\frac{2}{5}R_\nu }{1+\frac{4}{15}R_\nu} \zeta$ where $R_\nu=\rho_\nu/(\rho_\nu + \rho_\gamma)$ after the neutrino decoupling~\cite{Ma:1995ey}.
However, since we here consider the perturbations which reenter the horizon soon after the neutrino decoupling, we simply used the factor of $2/3$. Anyway this difference solely cause a factor modification on 
the curvature perturbation constraints.}
The detailed derivations of Eqs.~(\ref{g_potential}--\ref{theta1}) are described in the appendix \ref{app:decay_mode}. 
In the superhorizon limit ($k\eta \rightarrow 0$), it can be found that $\Phi \rightarrow \Phi_p$, $\Theta_0 \rightarrow \frac{\Phi_p}{2}$, and $\Theta_1 \rightarrow 0$. 
On the other hand, in the subhorizon limit ($k\eta \gg 1$), the gravitational potential decays $\Phi \rightarrow 0$ and the temperature perturbations oscillate as 
$\Theta_0 \rightarrow  -\frac{3}{2} \Phi_p \cos(k\eta /\sqrt{3})$ and $\Theta_1 \rightarrow  -\frac{\sqrt{3}}{2} \Phi_p \sin(k\eta /\sqrt{3})$. 
Here note that the amplitude of the oscillation of temperature perturbations $\bar{\Theta}^2=\Theta_0^2+3\Theta_1^2$ is given by $\frac{9}{4}\Phi_p^2$ and time-independent. 
Therefore this relation is often used in the calculation of CMB distortions for example. 
However, around the horizon scale, the decaying modes ($\sim \mathcal{O}(1/k\eta)$, $\mathcal{O}(1/(k\eta)^2)$) cannot be neglected and the amplitude of the temperature perturbations
are suppressed by factor 9 as $\bar{\Theta}^2\simeq\frac{\Phi_p^2}{4}$. Indeed this fact yields factor modifications for the constraints on the curvature perturbations.

\begin{figure}
	\centering
	\includegraphics[width=0.9\hsize]{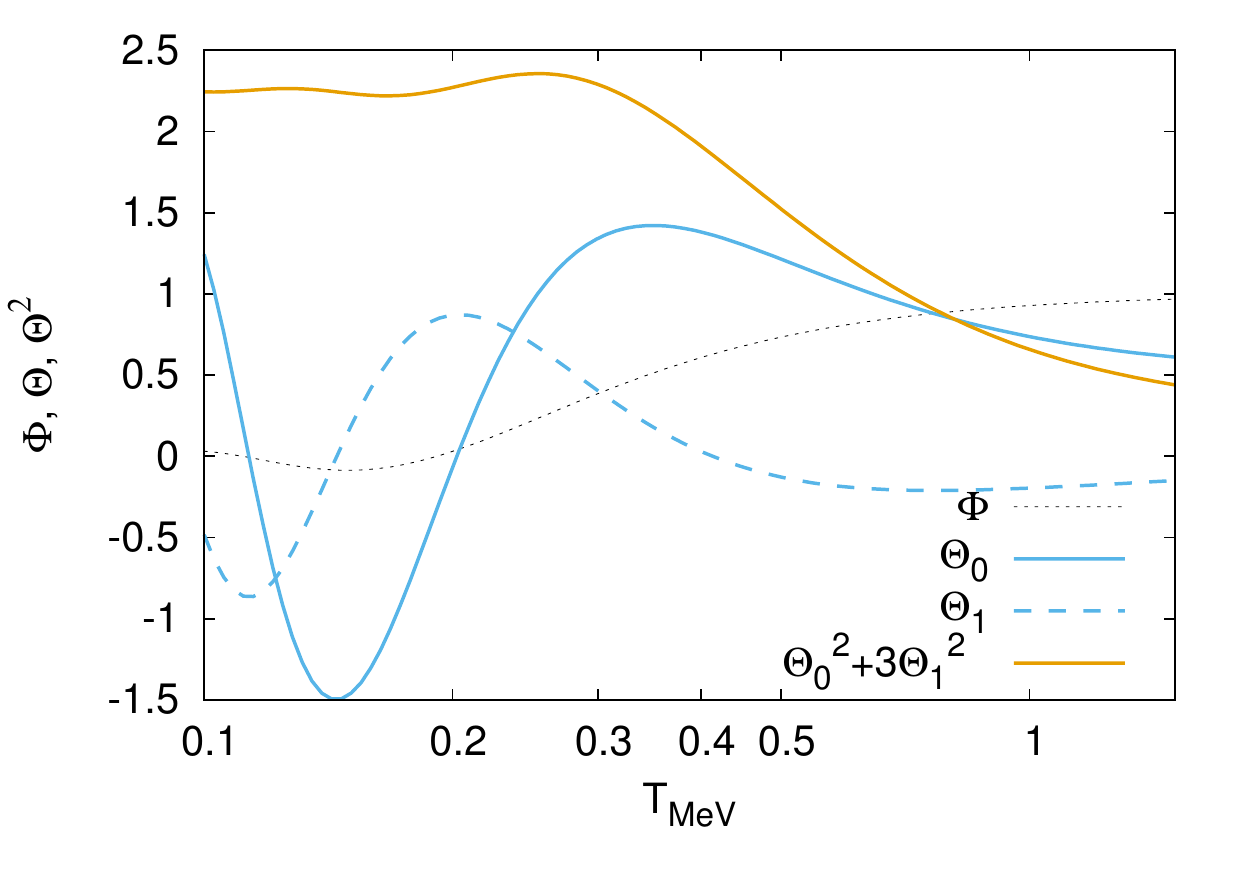}
	\caption{Plot of $\Phi, \Theta_0, \Theta_1,$ and $\Theta_0^2+3\Theta_1^2$ ($0.1\text{MeV} <T<1.5\mathrm{MeV}$), where we set $\Phi_p =1$. }
	\label{fig:theta}
\end{figure}

Now, in order to understand the temperature dependence of the perturbations visually, we consider the perturbation 
which enter the horizon soon after neutrinos are decoupled. Namely, assuming that neutrinos are decoupled at $T=1.5\,\mathrm{MeV}$ instantaneously, 
we focus on the mode which satisfies $k\eta=1$ at $T=1.5\,\mathrm{MeV}$. Indeed it is found that the concrete value of $k$ is around $10^5\,\mathrm{Mpc}^{-1}$  with use of the following relation
between time and temperature in the radiation dominated era~\cite{Mukhanov:2005}:
\begin{align}\label{time_temp}
	t_\mathrm{sec}\simeq1.39\kappa^{-1/2}\frac{1}{T_\mathrm{MeV}^2},
\end{align}
where $t_\text{sec}$ and $T_\text{MeV}$ are the cosmic time and temperature in units of sec and MeV, respectively, and $\kappa=\frac{\pi^2}{30}g_*=3.537\left(\frac{g_*}{10.75}\right)$ with $g_*$ being the effective degree of freedom (d.o.f). 
In that era, the conformal time satisfies $\eta \propto a \propto \sqrt{t}$ and therefore $k\eta \simeq \frac{1.5}{T_\mathrm{MeV}}$. 
Then now we can illustrate the solutions~(\ref{g_potential}--\ref{theta1}) and we plot them for $\Phi_p=1$ in Fig.~\ref{fig:theta}. 
From this figure, it can be found that $\Theta_0$ and $\Theta_1$ have not yet started oscillation around the freeze out temperature ($T\simeq 0.5\,\mathrm{MeV}$), and therefore the assumption that
neutrino perturbations are not yet smoothed off is justified.

\section{Calculation of neutron-proton ratio}
\label{n-p ratio}

In this section we discuss the calculation of the n-p ratio. 
In the following, we define  $X_n=\frac{n_n}{n_n+n_p}$, where $n_n$ and $n_p$ are the number densities of neutrons and protons, and call $X_n$ the n-p ratio.

\subsection{n-p ratio in the homegeneous universe}

In this subsection, we review the calculation of the n-p ratio without the perturbations briefly.
The n-p ratio is determined by the weak interaction among neutrons, protons, electrons (positrons), and neutrinos (antineutrinos). 
The time-evolution of $X_n$ is given by~\cite{Mukhanov:2005,Weinberg:2008},
\begin{align}\label{xn_fund}
	\dif{X_n(T)}{t} = -\lambda_{n \rightarrow p}X_n(T)+\lambda_{p\rightarrow n}(1-X_n(T)),
\end{align}
where $\lambda_{n\to p}$ and $\lambda_{p\to n}$ are the reaction rates of conversions from neutrons to protons and vice vasa.
The conversion from neutrons to protons is given by two processes, $n+\nu\to p+e^-$ and $n+e^+\to p+\bar{\nu}$, whose reaction rates are referred to as $\lambda_{n\nu}$ and $\lambda_{ne}$ respectively.
Namely $\lambda_{n\to p}=\lambda_{n\nu}+\lambda_{np}$. 
In fact there is a neutron-decay process ($n\to p+e^-+\bar{\nu}$) but the lifetime of neutron is so long as $\tau_n\sim886\,\mathrm{s}$ and it does not work until sufficient low temperature $\lesssim 0.05\,\mathrm{MeV}$. 
Therefore we will neglect the neutron decay and simply calculate the freeze-out value of the n-p ratio.
On the other hand, the reaction rate of the conversion from protons to neutrons is determined by $p+\bar{\nu}\to n+e^+$ and $p+e^-\to n+\nu$, neglecting the three-body process $p+e^-+\bar{\nu}\to n$. 
Accordingly the total reaction rate can be written as $\lambda_{p\to n}=\lambda_{p\nu}+\lambda_{pe}$.  

\begin{figure}[t]
	\centering		
	\includegraphics[width=0.9\hsize]{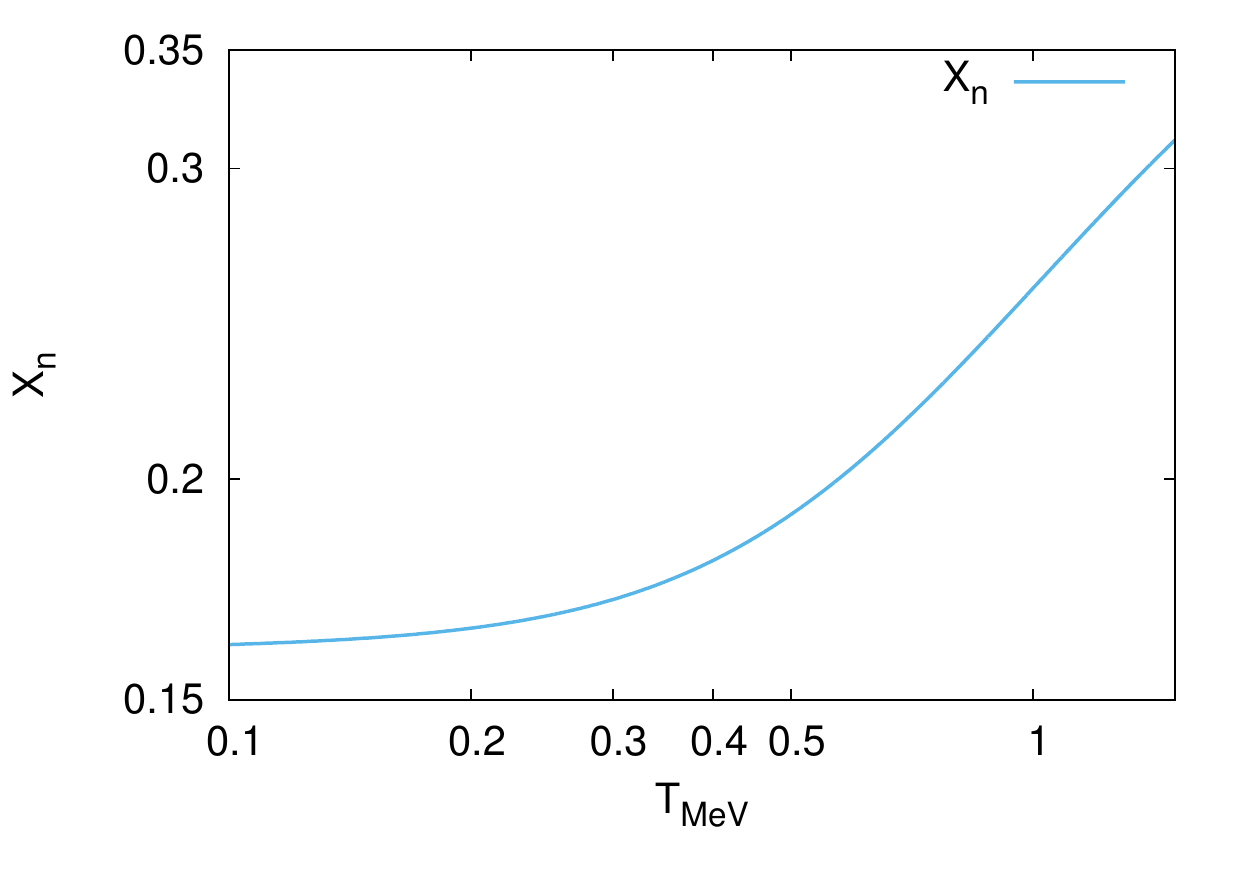}
	\caption{Evolution of $X_n$ without the perturbations ($0.1\,\mathrm{MeV}<T<1.5\,\mathrm{MeV}$). In this plot, it can be seen that the freeze out is occurred around $T\sim0.5\,\Mev$.}
	\label{fig:exact_no_fl}
\end{figure}

Each reaction rate can be calculated as,
\begin{align}
	\label{l_nnu}
	\lambda_{n\nu} &= \frac{1+3g_A^2}{2\pi^3}G_F^2  Q^5 J(1;\infty), \\
	\lambda_{ne} &= \frac{1+3g_A^2}{2\pi^3}G_F^2  Q^5 J(-\infty;-\frac{m_e}{Q}), \\
	\lambda_{p\nu} &= \frac{1+3g_A^2}{2\pi^3}G_F^2  Q^5 \hat J(1;\infty), \\
	\label{l_pe}
	\lambda_{pe} &= \frac{1+3g_A^2}{2\pi^3}G_F^2  Q^5 \hat J(-\infty;-\frac{m_e}{Q}),
\end{align}
where $g_A\simeq 1.26$ is the axial-vector coupling, $G_F\simeq 1.17\times 10^{-5}\,\Gev^{-2}$ is the Fermi coupling constant, and $Q= m_n -m_p \simeq 1.293\,\Mev$ is the mass difference between neutron and proton.
$J(a; b)$ and $\hat J (a; b)$ are defined as,
\begin{align}
	J(a;b)=&\int^b_a \sqrt{1-\frac{(m_e/Q)^2}{q^2}} \frac{q^2(q-1)^2\dd q}{(1+\ee^{\frac{Q}{T_\nu}(q-1)})(1+\ee^{-\frac{Q}{T}q})}, \\
	\hat J(a;b)=&\int^b_a \sqrt{1-\frac{(m_e/Q)^2}{q^2}} \frac{q^2(q+1)^2\dd q}{(1+\ee^{\frac{Q}{T_\nu}(q+1)})(1+\ee^{-\frac{Q}{T}q})}.
\end{align}
Here $T_\nu$ represents neutrino temperature, which is equal to photon temperature $T$ before the electron-positron annihilation.
Note that the denominators come from Fermi distribution function $\left(\frac{1}{1+\ee^{E/T}}\right)$ and the Pauli blocking factor $\left(1-\frac{1}{1+\ee^{E/T}}\right)$.
Using Eqs.~(\ref{time_temp}), (\ref{l_nnu})--(\ref{l_pe}), we can solve Eq. (\ref{xn_fund}) and plot the evolution of $X_n$ in Fig.~\ref{fig:exact_no_fl}, where it is seen that the n-p ratio is frozen around $T\sim0.5\,\Mev$.

\subsection{n-p ratio in the inhomogeneous universe}
\label{taking account of perturbations}

\begin{figure}[t]
	\centering
	\includegraphics[width=0.8\hsize]{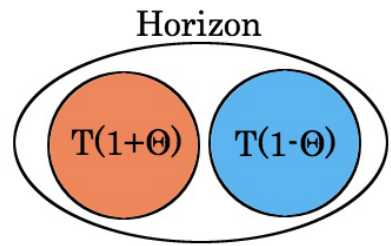}
	\caption{Schematic image of the two patches in the horizon, one of which has slightly high temperature $T(1+\Theta)$ and the other of which
	has slightly low temperature $T(1-\Theta)$.}
	\label{fig:patch}
\end{figure}

In this subsection, we discuss how we should take account of the effects of the perturbations. 
To make things simple, let us assume an illustrative situation that only two same comoving volume patches enter the horizon soon after the neutrino decoupling and
one has slightly high temperature $T(1+\Theta)$ and the other has 
slightly low temperature $T(1-\Theta)$.\footnote{Note that in this subsection $\Theta$ is a positive number representing the variance of
the temperature perturbations, while it will directly denote the perturbation value in each patch which becomes both positive and negative
in subsection~\ref{gaussian distribution}.
}
 The situation is schematically shown in Fig.~\ref{fig:patch}.
Of course the horizon will contain many patches well after the neutrino decoupling and we will extend the calculation to such a case in subsection~\ref{gaussian distribution}.

Let us calculate the reaction rate in the high temperature patch ($T(1+\Theta)$). In that patch, $J(a;b)$ and $\hat{J}(a;b)$ are modified to $J(a;b,\Theta)$ and $\hat{J}(a;b,\Theta)$, which are defined as, 
\begin{widetext}
	\begin{align}
		J(a;b,\Theta) &=\int^1_{-1} \frac{\dd\mu}{2} \int^b_a \sqrt{1-\frac{(m_e/Q)^2}{q^2}}
		\,\frac{q^2(q-1)^2}{(1+\ee^{\frac{Q}{T_\nu(1+\Theta)}(q-1)})(1+\ee^{-\frac{Q}{T(1+\Theta)}q})}\dd q, \\
		\hat{J}(a;b,\Theta)&=\int^1_{-1} \frac{\dd\mu}{2} \int^b_a \sqrt{1-\frac{(m_e/Q)^2}{q^2}}
		\,\frac{q^2(q+1)^2}{(1+\ee^{\frac{Q}{T_\nu(1+\Theta)}(q+1)})(1+\ee^{-\frac{Q}{T(1+\Theta)}q})} \dd q.
	\end{align}
\end{widetext}
Then, let $X_n^+$ denote the n-p ratio in the high temperature patch whose  evolution is determined by,
\begin{align}\label{xn+}
	\dif{X_n^+(T,\zeta)}{t} = &-\lambda_{n \rightarrow p}(\Theta)X_n^+(T,\zeta) \nonumber \\
	&+\lambda_{p\rightarrow n}(\Theta)(1-X_n^+(T,\zeta)),
\end{align}
where $\lambda_{n\rightarrow p}(\Theta) = \lambda_{n\nu}(\Theta) +\lambda_{ne}(\Theta) $ and $\lambda_{p\rightarrow n}(\Theta) = \lambda_{p\nu}(\Theta) +\lambda_{pe}(\Theta) $ are given by,
\begin{align}
	\label{l_nnu_fl}
	\lambda_{n\nu}(\Theta) &= \frac{1+3g_A^2}{2\pi^3}G_F^2  Q^5 J(1;\infty,\Theta),\\
	\label{l_ne_fl}
	\lambda_{ne}(\Theta) &= \frac{1+3g_A^2}{2\pi^3}G_F^2  Q^5 J(-\infty;-\frac{m_e}{Q},\Theta),\\
	\label{l_pnu_fl}
	\lambda_{p\nu}(\Theta) &= \frac{1+3g_A^2}{2\pi^3}G_F^2  Q^5 \hat J(1;\infty,\Theta),\\
	\label{l_pe_fl}
	\lambda_{pe}(\Theta) &= \frac{1+3g_A^2}{2\pi^3}G_F^2  Q^5 \hat J(-\infty;-\frac{m_e}{Q},\Theta).
\end{align}
With use of Eq.~(\ref{time_temp}), we can solve Eq.~(\ref{xn+}) and derive the evolution of $X_n^+$.

On the other hand, in the low temperature patch we only have to change the sign of $\Theta$ ($\Theta \rightarrow -\Theta$). Then the evolution of n-p ratio in that patch ($X_n^-$) are described as,
\begin{align}\label{xn-}
	\dif{X_n^-(T,\zeta)}{t} = &-\lambda_{n \rightarrow p}(-\Theta)X_n^-(T,\zeta) \nonumber \\
	&+\lambda_{p\rightarrow n}(-\Theta)(1-X_n^-(T,\zeta)).
\end{align}
Similarly to $X_n^+$, we can obtain the temperature dependence of $X_n^-$ from this equation with use of Eq.~(\ref{time_temp}).

Finally, we have to mention how we should average the two n-p ratio, $X_n^+$ and $X_n^-$. 
If the perturbations are adiabatic ($\Theta_0 = \frac{1}{3}\delta_b = \frac{1}{3}\delta n_b$), the two patches have different number density of baryons. 
Since $X_n$ is just a ratio of neutrons to baryons, we have to weight it by the number of baryons included in each patch.
Here we have assumed that two patches have the same comoving volume ($V_0$) at the horizon cross.
Therefore the physical volumes of the two patches are given by $V_0 a^3 (1+3\Phi^*)$ (hot) and $V_0 a^3 (1-3\Phi^*)$ (cold), 
where ${}^*$ means the value at the horizon cross .
For example, the baryon number ($N_b$) included in the high temperature patch ($T(1+\Theta)$) is given by
\begin{align}
	N_b &= (1+\delta_b^*) \frac{n_b^*}{a^3}  a^3 (1+3\Phi^*) V_0 \nonumber \\
		&\simeq (1+3\Theta_0^*) n_b^* (1+6\Theta_0^*) V_0 \nonumber \\
		&\simeq (1+9\Theta_0^*) n_b^* V_0,
		\label{newtonian_patch}
\end{align}
where $n_b$ is the averaged baryon number density and we use the fact $\Phi^* \simeq 2\Theta_0^*$.

To check the validity of Eq.~(\ref{newtonian_patch}), let us rederive
the weight factor
in the spatially flat gauge as well as 
the conformal Newtonian gauge.
That is because it is trivial how to take spatial averages in the spatially flat gauge since
there is no metric perturbation and no difference between a comoving and physical volume.
Between these gauges,
one can derive the relation $\Theta^{*\,\mathrm{flat}}=3\,\Theta^{*\,\mathrm{Newtonian}}$, using two expressions of
the gauge invariant curvature perturbation $\zeta=\frac{1}{3}\frac{\delta\rho^\mathrm{flat}}{\rho+P}
=\Phi^\mathrm{Newtonian}+\frac{1}{3}\frac{\delta\rho^\mathrm{Newtonian}}{\rho+P}$.
Then, the baryon number included in the high temperature patch is given by,
\begin{align}
N_b &= (1+\delta_b^{*\mathrm{flat}}) \frac{n_b^*}{a^3}  a^3  V_0 \nonumber \\
&\simeq (1+3\Theta_0^{*\mathrm{flat}}) n_b^* V_0 \nonumber \\
&\simeq (1+9\Theta_0^{*\mathrm{Newtonian}}) n_b^* V_0,
\label{flat_patch}
\end{align}
which is indeed consistent with Eq.~(\ref{newtonian_patch}).

Then the averaged $X_n$ is given by
\begin{align}
	\bar{X}_n(\sigma) \!=\! \frac{1}{2}\left((1\!+\!9\Theta_0^*)X_n^+(T\!\to\!0) \!+\! (1\!-\!9\Theta_0^*)X_n^-(T\!\to\!0)\right),
	\label{sum_patch}
\end{align}
where $\sigma$ represents the amplitude of the perturbations, $\sigma^2=\braket{\zeta^2}$.

After the freeze-out of $X_n$, neutrons gradually convert to protons due to their decay process until $^4$He synthesis.
Therefore, the final $^4$He abundance with perturbations can be written as $Y_p(\sigma)=2\bar{X}_n(\sigma)D$ with the decay factor $D$.
The decay factor can be approximated as \cite{Mukhanov:2005},
\begin{align}\label{D}
D\simeq \mathrm{exp}\left[-\frac{269(1-0.07(N_\nu-3)-0.06\,\mathrm{ln}\,\eta_{10})}{\tau_n}\right],
\end{align}
where $\tau_n\simeq 886 \,\mathrm{s}$ is the life time of a free neutron, $N_\nu$ is the number of neutrino species,
and $\eta_{10}=10^{10}\times n_b/n_\gamma$ represents the baryon-photon ratio.
Without perturbations, the Planck consistent decay factor can be easily evaluated by $D=Y_p^\mathrm{CMB}/2\bar{X}_n(\sigma=0)$ 
where $Y_p^\mathrm{CMB}=0.24668$ is the central value of BBN prediction with Planck's best fit parameters~\cite{Ade:2015xua}.
However, if there are large density perturbations, their decay due to the Silk damping causes energy injection and increases
the photon number density after BBN but before the last scattering surface. 
Therefore the baryon-photon ratio during the BBN era should be larger than that evaluated by the CMB observation,
and that modification is given by~\cite{Nakama:2014vla},
\begin{align}
	\frac{\Delta\eta_{10}}{\eta_{10}}=\frac{\eta_{10}^\mathrm{BBN}-\eta_{10}^\mathrm{CMB}}{\eta_{10}^\mathrm{BBN}}
	\simeq\frac{3}{4}\times2.3\braket{\zeta^2}.
\end{align} 
Hence it possibly changes the decay factor, though the n-p ratio itself does not depend on the baryon-photon ratio directly.

From Eq.~(\ref{D}), the modification of the decay factor is estimated as,
\begin{align}
&D+\Delta D \nonumber\\
&\simeq \mathrm{exp}\left[-\frac{269(1-0.07(N_\nu-3)-0.06\,\mathrm{ln}\,(\eta_{10}+\Delta \eta_{10}))}{\tau_n}\right],
\end{align}
and it gives,
\begin{align}
\frac{\Delta D}{D} \simeq 0.018\times\frac{\Delta \eta_{10}}{\eta_{10}}\simeq0.031\braket{\zeta^2},
\end{align}
with concrete parameter values.
However it is much smaller than the modification of n-p ratio itself $\Delta\bar{X}_n/\bar{X}_n\simeq-2.2\braket{\zeta^2}$ as
we will see in Fig.~\ref{fig:sigma_n-p}. Therefore the modification of the decay factor can be neglected.\footnote{Note that 
Nakama et al.~\cite{Nakama:2014vla} have given the constraints on the curvature perturbation by the modification of $\eta_{10}$
with use of the deutron abundance, which is sensitive to the value of $\eta_{10}$. However the constraints are weaker 
($\Delta_\mathcal{R}^2\lesssim0.06$) and independent of our constraints with the n-p ratio.}

With use of the decay factor $D=Y_p^\mathrm{CMB}/2\bar{X}_n(\sigma=0)$, we can obtain the constraints on 
the freeze-out value $\bar{X}_n(\sigma)$ with perturbations, so that the $^4$He abundance to be consistent with 
the observational value $Y_p^\mathrm{obs}\pm\Delta Y_p^\mathrm{obs}=0.2449\pm0.0080~(2\sigma)$~\cite{Aver:2015iza}, as,

\begin{align}\label{def of constraints}
	&\frac{Y_p^\mathrm{obs}-\Delta Y_p^\mathrm{obs}}{2D}<\bar{X}_n(\sigma)<\frac{Y_p^\mathrm{obs}+\Delta Y_p^\mathrm{obs}}{2D}, \nonumber \\
	\Leftrightarrow& \frac{Y_p^\mathrm{obs}-\Delta Y_p^\mathrm{obs}}{Y_p^\mathrm{CMB}}<\frac{\bar{X}_n(\sigma)}{\bar{X}_n(\sigma=0)}<\frac{Y_p^\mathrm{obs}+\Delta Y_p^\mathrm{obs}}{Y_p^\mathrm{CMB}}.
\end{align}
This eventually gives the constraint on the primordial curvature perturbations.

\subsection{Approximations of reaction rates}
\label{approximations of reaction rates}

In the previous subsection, we showed how we can calculate the averaged $X_n$ exactly. 
However, for the calculation over many patches in the next subsection, we would like to use some approximated forms of the reaction rates to avoid time-consuming numerical integrations. 
Though we postpone the detail derivation to appendix~\ref{app:approximation}, we can obtain the following approximations under the assumptions that temperature is sufficiently higher than the mass difference $T\gg Q$ and the Pauli-blocking factors are negligible:
\begin{align}
	\label{l_nnu_app_wop}
	\lambda_{n\nu} &\simeq 1.63 \left( \frac{T_\nu}{Q} \right)^3 \left( \frac{T_\nu}{Q} + 0.25 \right)^2 \mathrm{s}^{-1}, \\
	\lambda_{ne} &\simeq 1.63 \left( \frac{T}{Q} \right)^3 \left( \frac{T}{Q} + 0.25 \right)^2 \mathrm{s}^{-1}, \\
	\lambda_{p\nu} &\simeq 1.63\,\ee^{-\frac{Q}{T_\nu}}\left( \frac{T_\nu}{Q} \right)^3 \left( \frac{T_\nu}{Q} + 0.25 \right)^2 \mathrm{s}^{-1}, \\
	\label{l_pe_app_wop}
	\lambda_{pe} &\simeq 1.63\,\ee^{-\frac{Q}{T}}\left( \frac{T}{Q} \right)^3 \left( \frac{T}{Q} + 0.25 \right)^2 \mathrm{s}^{-1}.
\end{align}

If we include the perturbation, the reaction rates reads (see appendix~\ref{app:taylor_ex} for detail derivations), up to $\mathcal{O}(\Theta^4)$,
\begin{widetext}
	\begin{align}
		\label{l_nnu_app}
		\lambda_{n\nu}(\Theta) &\simeq \lambda_{ne}(\Theta) \nonumber \\
		 &\simeq 1.63\left( \frac{T}{Q} \right)^3(a+(3a+b)\Theta_0 + (3a+3b+c) \bar{\Theta}^2) \,\mathrm{s}^{-1}, \\
		\label{l_pnu_app}
		\lambda_{p\nu}(\Theta)&\simeq \lambda_{pe}(\Theta) \nonumber \\
		&\simeq 1.63\,\ee^{-\frac{Q}{T}}\left( \frac{T}{Q} \right)^3
		\left(a + \left(3a+b+a\frac{Q}{T}\right)\Theta_0 + \left(a\left[\frac{1}{2}\left(\frac{Q}{T}\right)^2-\frac{Q}{T}\right]+3a+3b+c+\frac{Q}{T}(3a+b)\right)\bar{\Theta}^2\right) \,\mathrm{s}^{-1},
	\end{align}
\end{widetext}
where we have assumed $T=T_\nu$ and $a, b,$ and $c$ are given by,
\begin{align}
	a&= \left(\frac{T}{Q} \right)^2 + \frac{T}{2Q} + \frac{1}{16},\\
	b&= 2\left( \frac{T}{Q} \right)^2 + \frac{T}{2Q}, \\
	c&= \left( \frac{T}{Q} \right)^2. 
\end{align}

\begin{figure}
	\centering
	\includegraphics[width=0.9\hsize]{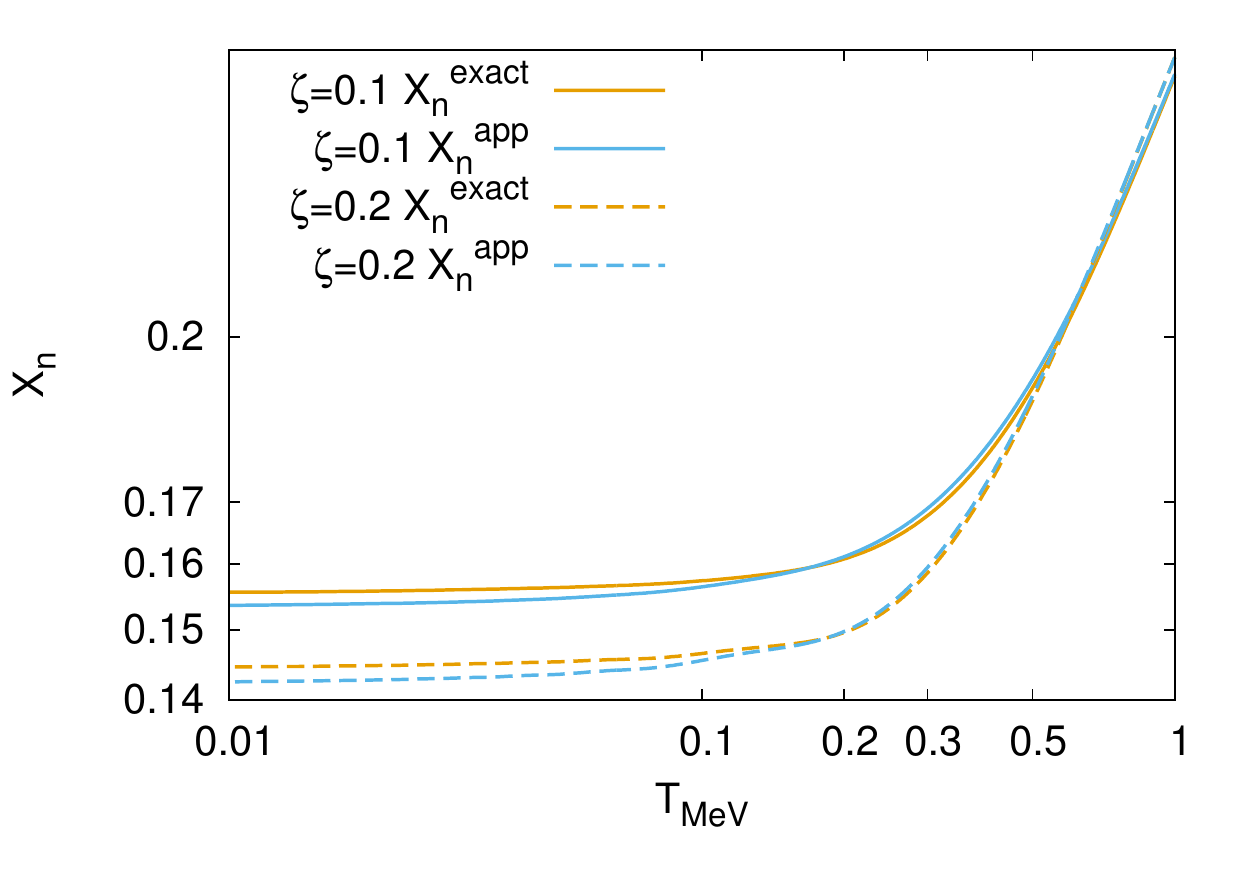}
	\caption{Evolution of $X_n^{\mathrm{exact}}$ (orange) and $X_n^{\mathrm{app}}$ (blue) for $\zeta = 0.1$ (dotted) and $0.2$ (solid). }
	\label{fig:xn_exact_app_e}
\end{figure}

Now let us evaluate the precision of the approximated forms.
Since these forms use the high-temperature limit as mentioned, they are expected not to be valid at low temperature. 
Indeed as shown in Fig.~\ref{fig:xn_exact_app_e} $X_n^\mathrm{app}$ which is obtained by the approximated reaction rates~(\ref{l_nnu_app}--\ref{l_pnu_app}) starts to deviate from the exact $X_n^\mathrm{exact}$ around $T\sim0.2\,\Mev$.

However it is expected that the dynamics of $X_n$ after the freeze-out will not depend on the perturbations so much.
Therefore the final value of $X_n$ could be obtained by multiplying $X_n$ at some (not-so-low) temperature by some numerical factor which is independent of the amplitude of perturbations.

\begin{figure}
	\centering
	\includegraphics[width=0.9\hsize]{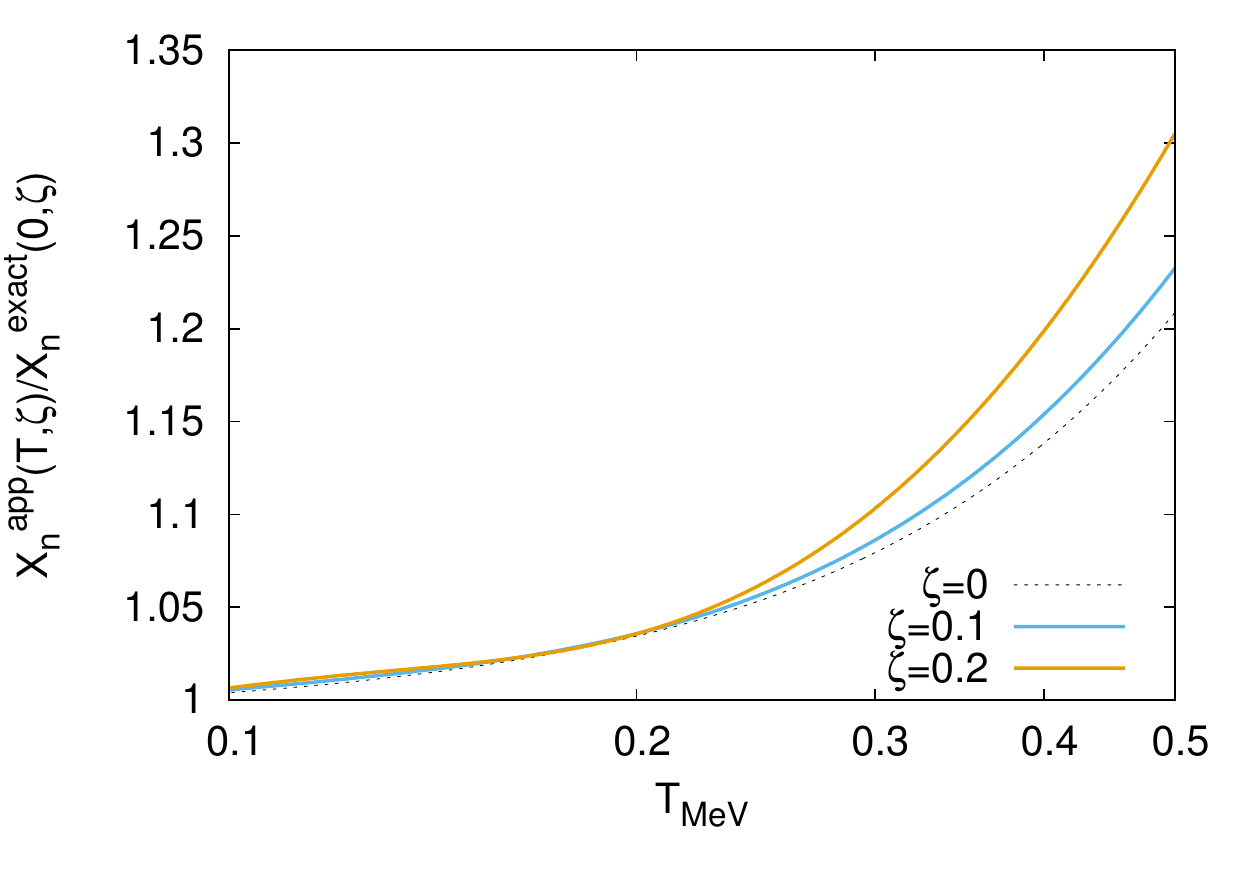}
	\caption{Ratio of $X_n^{\mathrm{exact}}(0,\zeta)$ to $X_n^{\mathrm{app}}(T,\zeta)$ with $\zeta = 0,\,0.1,$ and $0.2$. }
	\label{fig:freeze_sigma}
\end{figure}

In Fig.~\ref{fig:freeze_sigma}, we plot the ratio of $X_n^\mathrm{app}(T,\zeta)$ to $X_n^\mathrm{exact}(0,\zeta)$, which shows that the ratio does not depend on the value of perturbations and hence justifies the above expectation. 
It means that the evolution of $X_n$ at low temperature $T\lesssim 0.2\,\Mev$ can be represented by the simple rescaling with some numerical factor irrespectively of the perturbations. 
In this paper, we approximate $X_n^\mathrm{exact}(T\to0)$ by $X_n^\mathrm{app}(T=0.2\,\mathrm{MeV})/1.034$ hereafter.

\subsection{Gaussian distribution}
\label{gaussian distribution}

In this subsection we consider more than two patches and obtain the concrete constraint on the curvature perturbations.
With the assumption that the curvature perturbations $\zeta$ follow the Gauss distribution, various quantities can be averaged over the Gauss distribution.
In the following we define $\sigma^2$ as the variance of the curvature perturbations $\braket{\zeta^2}$. 

We already see how the n-p ratio should be calculated in each patch in subsection~\ref{taking account of perturbations}. For example, in the high temperature patch we can derive the freeze out value of the n-p ratio using Eq.(\ref{xn+}).
Then the n-p ratio in each patch should be averaged over the probability distribution of $\Theta$.
As mentioned in subsection~\ref{taking account of perturbations}, the weight factor is given by $1+9\Theta_0^*$ with $\Theta_0^*=\zeta/3$
being the initial condition of $\Theta_0$.\footnote{

Note that the positive $\zeta$ corresponds to the hot patch and the negative $\zeta$ corresponds to the cold patch.
$\Theta_0^*$ takes both positive and negative values now}.
Thus, the averaged n-p ratio $\bar{X}_n^\text{ave}$ is given by,
\begin{align}
	\bar{X}^{\mathrm{ave}}_n(T\to0,\sigma) =&\int^\infty_{-\infty} \dd\zeta \frac{1}{\sqrt{2\pi}\sigma}\ee^{\frac{-\zeta^2}{2\sigma^2} }(1+9\Theta_0^*)\frac{X_n(0.2,\zeta)}{1.034},
\end{align}
where we use the approximated reaction rates (\ref{l_nnu_app}--\ref{l_pnu_app}) and derive $X_n(T_\Mev=0.2,\zeta)/1.034$ as the final freeze-out value of the n-p ratio as discussed in subsection~\ref{approximations of reaction rates}.
In the actual calculations, we have used the Gaussian distribution truncated at 3$\sigma$ as,
\begin{align}
	P_{G}&=\frac{1}{\sqrt{2\pi}\sigma}\ee^{\frac{-\zeta^2}{2\sigma^2}}, \quad -\infty<\zeta<\infty, \nonumber \\[5pt]
	\hspace{-5pt}
	\to P_{G^\prime}&=
	\begin{cases}
		\displaystyle
		\!\left(\!\int^{3\sigma}_{-3\sigma}\dd\zeta\frac{1}{\sqrt{2\pi}\sigma}\ee^{\frac{-\zeta^2}{2\sigma^2}}\!\right)^{-1}\!
		\frac{1}{\sqrt{2\pi}\sigma}\ee^{\frac{-\zeta^2}{2\sigma^2}}, & |\zeta|\le3\sigma, \\[10pt]
		\displaystyle
		0, & \text{otherwise},
	\end{cases}
\end{align}
to avoid including high-$\Theta$ values since we have used $\frac{Q}{T}\Theta$ expansions and the results for high $\Theta$ lose reliability.

\begin{figure}
	\centering
	\includegraphics[width=0.9\hsize]{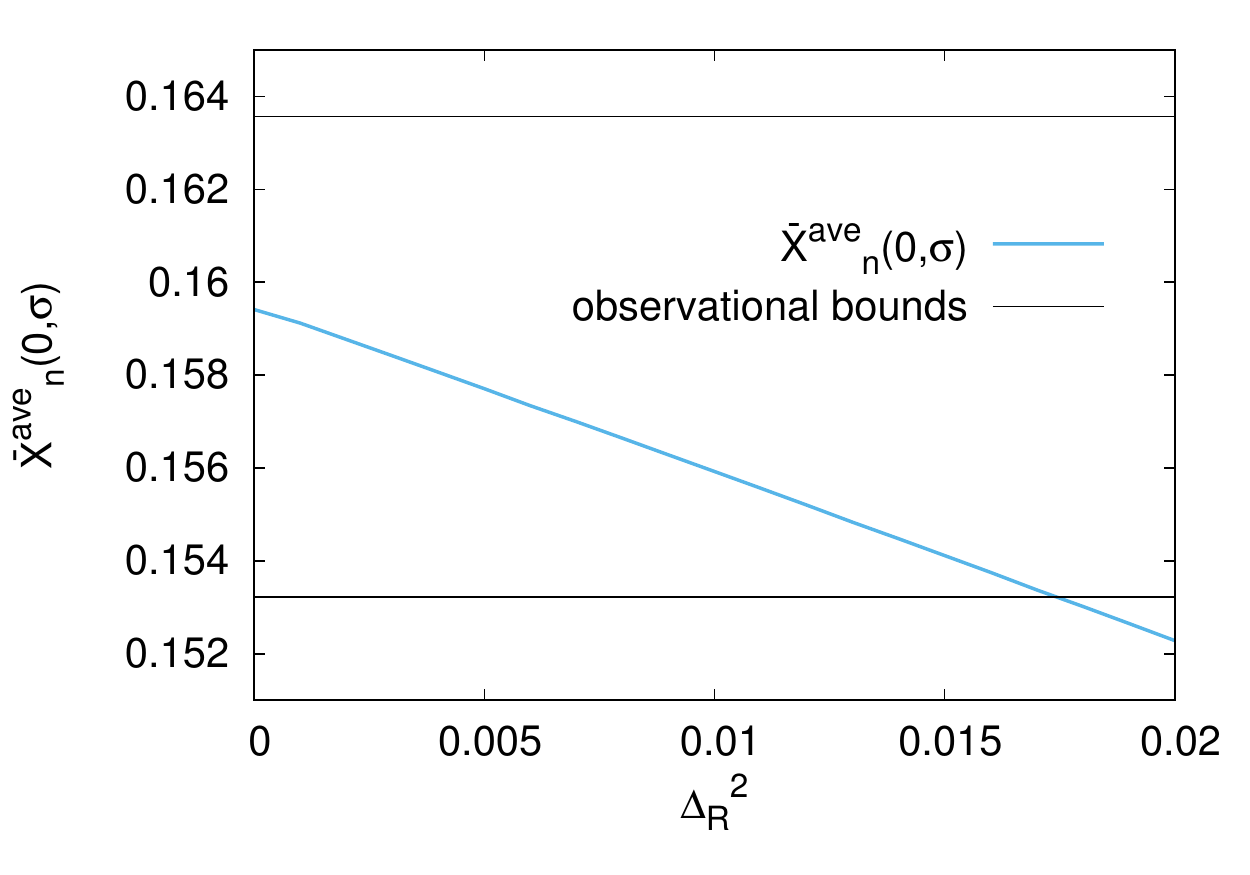}
	\caption{Plot of $\bar{X}_n^{\mathrm{ave}}(0,\sigma)$ (blue solid) and the latest observational bounds (black solid)~\cite{Aver:2015iza} .}
	\label{fig:sigma_n-p}
\end{figure}

We show the results as the blue solid line in Fig.~\ref{fig:sigma_n-p}. 
We also plot the observational bounds $\frac{Y_p^\mathrm{obs}\pm\Delta Y_p^\mathrm{obs}}{Y_p^\mathrm{CMB}}\bar{X}_n^\mathrm{ave}(0,0)$ as black solid lines with use of the latest value $Y_p^\mathrm{obs}\pm\Delta Y_p^\mathrm{obs}=0.2449\pm0.0080~(2\sigma)$~\cite{Aver:2015iza} .
From this figure we obtain the constraint on $\Delta_\mathcal{R}^2$ as,
\begin{align}
	\Delta_\mathcal{R}^2\lesssim 0.018, \quad k\sim10^5\,\mathrm{Mpc}^{-1}.
\end{align}
This is the main conclusion of this paper.

For modes with slightly larger wavelengths $10^4\,\mathrm{Mpc}\lesssim k<10^5\,\mathrm{Mpc}^{-1}$ which enter the horizon somewhat after the neutrino decoupling, we have checked that the constraint is weakened since those modes do not have much time between the horizon cross and the freeze-out.

Finally, we consider why the perturbations make the freeze out value of the n-p ratio decrease.
It can be interpreted as the effect of the existence of local patches.
The reaction rates are determined by the temperature in each patch. 
Therefore in the high temperature patch the reaction rates are enhanced and the freeze-out is delayed, which leads to  the smaller final freeze-out value. On the other hand, in the low temperature patch, the freeze-out is advanced and the final freeze-out value becomes larger. 
Here recall that these values should be averaged with a weight factor $1+9\Theta^*$.
Namely the smaller freeze out value in the high temperature patch is weighted by the heavier weight $1+9\Theta^*$, while the larger freeze out value in the low temperature patch is weighted by the lighter weight $1-9\Theta^*$. 
Therefore this effect works to decrease the averaged freeze-out value.
From our results (Fig.~\ref{fig:sigma_n-p}), it can be found that such decreasing effects give a dominant contribution.

\section{Conclusions}
\label{conclusions}

In this paper we revisit the modification of the neutron-proton ratio with large density perturbations on small scales $10^4\,\mathrm{Mpc}^{-1}\lesssim  k\lesssim 10^5\,\mathrm{Mpc}^{-1}$. 
Previously two groups of Jeong et al.~\cite{Jeong:2014gna} and Nakama et al.~\cite{Nakama:2014vla} considered the constraints on the primordial perturbations from BBN. 
Jeong et al. gave the constraint of $\Delta_\mathcal{R}^2\lesssim 0.007$ from the increase of the $^4$He abundance which is directly related with the n-p ratio, while the constraint by Nakama et al. are $\Delta_\mathcal{R}^2\lesssim 0.06$ from the modification of the baryon-photon ratio. 
Especially related with Jeong et al., we reconsider 
the distribution of electrons and neutrinos and the behavior of the perturbations around the horizon cross.
In fact, not only electrons but also neutrinos do evolve conjointly with baryon-photon fluid between the horizon reenter and the freeze out of the n-p ratio. Therefore the universe can be divided into
local patches which follow different thermal distributions. The existence of these local patches rather decrease the n-p ratio since they effectively delay the freeze out. Then our resultant constraint is 
$\Delta_\mathcal{R}^2\lesssim0.018$ with use of the latest observed value of the ${}^4$He abundance $Y_p^\mathrm{obs}=0.2449\pm0.0080\, (2\sigma)$~\cite{Aver:2015iza} .

\begin{acknowledgements}
We would like to thank Tomohiro Nakama and Atsuhisa Ota for meaningful discussions.
This work is supported by MEXT KAKENHI Grant Number 15H05889 (M. K.), JSPS KAKENHI Grant Number 25400248 (M. K.) and also by the World Premier International Research Center Initiative (WPI), MEXT, Japan.
Y. T. is supported by JSPS Research Fellowship for Young Scientists. K. I. is supported by Advanced Leading Graduate Course for Photon Science.
\end{acknowledgements}

\appendix

\section{Derivation of $\Phi, \Theta_0$, and $\Theta_1$}
\label{app:decay_mode}

In this appendix, we derive Eqs.~(\ref{g_potential}), (\ref{theta0}) and (\ref{theta1}). 
In the radiation dominated era, the time evolution of $\Theta_0$, $\Theta_1$ and $\Phi$ are described as~\cite{Dodelson:2003},
\begin{align}
	\label{app:theta0_eq}
	\dot{\Theta}_0+k\Theta_1 &= -\dot{\Phi},\\
	\label{app:theta1_eq}
	\dot{\Theta}_1-\frac{k}{3}\Theta_0&=\frac{-3}{k}\Phi, \\
	\label{app:phi_eq}
	\Phi &=\frac{6a^2H^2}{k^2}\left[\Theta_0+\frac{3aH}{k}\Theta_1\right],
\end{align}
where the overdots represent derivatives with respect to conformal time $\eta$.
Note that in the radiation dominated era the horizon scale is given by $aH=1/\eta$.
Using Eq.~(\ref{app:phi_eq}) and eliminating $\Theta_0$ in Eqs.~(\ref{app:theta0_eq}) and (\ref{app:theta1_eq}), we can obtain,
\begin{align}
	\label{app:theta1_a}
	-\frac{3}{k\eta}\dot{\Theta}_1 + k\Theta_1\left[ 1+\frac{3}{k^2\eta^2} \right] &=-\dot{\Phi} \left[1+\frac{k^2\eta^2}{6} \right] - \Phi\frac{k^2\eta}{3} ,\\
	\label{app:theta1_b}
	\dot\Theta_1+\frac{1}{\eta}\Theta_1 &= \frac{-k}{3}\Phi\left[1-\frac{k^2\eta^2}{6}\right]~.
\end{align}
Using Eq.~(\ref{app:theta1_b}) and eliminating $\dot\Theta_1$ in Eq.~(\ref{app:theta1_a}), we obtain,
\begin{align}
	\label{app:phi_a}
	\dot\Phi +\frac{1}{\eta}\Phi = \frac{-6}{k\eta^2}\Theta_1.
\end{align}
From Eqs.~(\ref{app:theta1_b}) and (\ref{app:phi_a}), we can get the following second-order equation:
\begin{align}
	\label{app:phi_b}
	\ddot\Phi +\frac{4}{\eta}\dot\Phi + \frac{k^2}{3} \Phi =0.
\end{align}
Eq.~(\ref{app:phi_b}) can be solved analytically by defining $u\equiv \Phi \eta$. Then Eq.~(\ref{app:phi_b}) reads,
\begin{align}
	\ddot u + \frac{2}{\eta} \dot u + \left( \frac{k^2}{3} -\frac{2}{\eta^2}\right) u =0.
\end{align}
This is the spherical Bessel equation of order 1 and the solutions are given by the spherical Bessel function $j_1(k\eta/\sqrt{3})$ and the spherical Neumann function $n_1(k\eta/\sqrt{3})$. 
Since the latter one blows up as $\eta$ gets very small, $n_1(k\eta/\sqrt{3})$ does not satisfy the initial condition. 
Therefore, we find $u\propto j_1(k\eta/\sqrt{3})$. 
The spherical Bessel function of order 1 can be expressed in terms of trigonometric functions as,
\begin{align}
	\label{app:phi_final}
	\Phi = -3\Phi_p\left(\left(\frac{\sqrt{3}}{k\eta}\right)^2\cos\frac{k\eta}{\sqrt{3}}-\left(\frac{\sqrt{3}}{k\eta}\right)^3\sin\frac{k\eta}{\sqrt{3}}\right),
\end{align}
where $\Phi_p$ is the primordial value of $\Phi$. We can easily confirm that $\Phi \rightarrow \Phi_p$  in the $\eta\rightarrow 0$ limit.

Substituting Eq.~(\ref{app:phi_final}) into Eqs.~(\ref{app:phi_eq}) and (\ref{app:phi_a}), we can easily get the solutions for $\Theta_0$ and $\Theta_1$ as,
\begin{align}
	\Theta_0&=-\frac{3}{2}\Phi_p\left(\cos\frac{k\eta}{\sqrt{3}}-\frac{\sqrt{3}}{k\eta}\sin\frac{k\eta}{\sqrt{3}}\right)-\frac{3}{k\eta}\Theta_1, \\
	\Theta_1&\!=\!-\frac{\sqrt{3}}{2}\Phi_p\!\left(\!\sin\frac{k\eta}{\sqrt{3}}\!+\!2\frac{\sqrt{3}}{k\eta}\cos\frac{k\eta}{\sqrt{3}}\!-\!2\!\left(\!\frac{\sqrt{3}}{k\eta}\!\right)^2\!\sin\frac{k\eta}{\sqrt{3}}\!\right)\!,
\end{align}
where we use the fact $aH=1/\eta$ in the radiation dominated era.

\section{The derivation of the approximation forms for the reaction rates}
\label{app:approximation}

In this appendix, we derive Eqs.~(\ref{l_nnu_app_wop})--(\ref{l_pe_app_wop}).
First, let us consider $\lambda_{n\nu}$:
\begin{align}
	\label{app:l_nnu}
	\lambda_{n\nu}= \frac{1+3g_A^2}{2\pi^3}G_F^2  Q^5 J(1;\infty).
\end{align}
To approximate $J(1;\infty)$, we use the fact that $m_e/Q \simeq 0.15$ with respect to which the reaction rate can be expanded. 
Neglecting the Pauli blocking factor, we can obtain,
\begin{align}
	\hspace{-5pt}
	 J(1;\infty)&=\int^\infty_1\sqrt{1-\frac{(m_e/Q)^2}{q^2}}\frac{q^2(q-1)^2\dd q}{(1+\ee^{\frac{Q}{T_\nu}(q-1)})(1+\ee^{-\frac{Q}{T}q})} \nonumber \\
 	&\simeq
 	\label{app:l_nnu_int_i}
  	\int^\infty_1 \left( 1-\frac{(m_e/Q)^2}{2q^2} \right)\frac{q^2(q-1)^2\dd q}{1+\ee^{\frac{Q}{T_\nu}(q-1)}}  \\
  	& = 
	\label{app:l_nnu_int}
   	\int^\infty_0 \left( 1-\frac{(m_e/Q)^2}{2(q+1)^2} \right)\frac{q^2(q+1)^2\dd q}{1+\ee^{\frac{Q}{T_\nu}q}} \\
	\label{app:l_nnu_int_f}
  	& = \int^\infty_0 \frac{q^4+2q^3+q^2\left( 1-\frac{1}{2} \left( \frac{m_e}{Q} \right) ^2 \right) }{1+\ee^{\frac{Q}{T_\nu}q}}\dd q.
\end{align}
Executing the integration, we find,
\begin{align} 
	\label{app:J(1:infty)}
  	J(1:\infty) \simeq& \frac{45\zeta(5)}{2} \left ( \frac{T_\nu}{Q} \right)^5 +\frac{7\pi^4}{60}  \left( \frac{T_\nu}{Q} \right)^4\nonumber \\
  	&+\frac{3\zeta(3)}{2}  \left( 1-\frac{1}{2} \left( \frac{m_e}{Q} \right) ^2 \right)  \left( \frac{T_\nu}{Q} \right)^3,
\end{align}
where $\zeta$ is the Riemann zeta function ($\zeta(3)\simeq 1.202$ and $\zeta(5) \simeq 1.037$).
Substituting Eq.~(\ref{app:J(1:infty)}) and the concrete values of $G_F,Q,$ and $g_A$ into Eq.~(\ref{app:l_nnu}), we can obtain the last form as,
\begin{align}
	\label{app:l_nnu_rough}
	\lambda_{n\nu} \simeq 1.63 \left( \frac{T_\nu}{Q} \right)^3 \left( \frac{T_\nu}{Q} + 0.25 \right)^2 \mathrm{s}^{-1}.
\end{align} 

Next, let us consider $\lambda_{ne}$:
\begin{align}
	\lambda_{ne} &= \frac{1+3g_A^2}{2\pi^3}G_F^2  Q^5 J(-\infty;-\frac{m_e}{Q}).
\end{align}
Similarly to $\lambda_{n\nu}$, we obtain,
\begin{align}
	\label{app:l_ne_int}
	J(-\infty;-\frac{m_e}{Q})=&\int^{-\frac{m_e}{Q}}_{-\infty}\sqrt{1-\frac{(m_e/Q)^2}{q^2}} \nonumber \\
	&\times\frac{q^2(q-1)^2\dd q}{(1+\ee^{\frac{Q}{T_\nu}(q-1)})(1+\ee^{-\frac{Q}{T}q})} \nonumber \\
 	\simeq&\int^\infty_{\frac{m_e}{Q}} \left( 1-\frac{(m_e/Q)^2}{2q^2}\right) \frac{q^2(q+1)^2\dd q}{1+\ee^{\frac{Q}{T}q}}.
\end{align}
Note that this expansion of the square root is valid only if the dominant contribution to the integral comes from large $q$ ($q\gg1$).
Comparing Eqs.~(\ref{app:l_ne_int}) and (\ref{app:l_nnu_int}), we find that if the contribution of large $q$ is dominant 
these two equations become nearly equal, that is, $\left( 1-\frac{(m_e/Q)^2}{2(q+1)^2}\right) \simeq  \left( 1-\frac{(m_e/Q)^2}{2q^2}\right)$ when $q$ is large.
The dominant contribution to the integral is determined by the factor $1/(1+\ee^{\frac{Q}{T}q})$ and large $q$ gives a dominant contribution for $Q/T<1$.
Thus we can derive the approximation form as,
\begin{align}
	\lambda_{ne} \simeq 1.63 \left( \frac{T}{Q} \right)^3 \left( \frac{T}{Q} + 0.25 \right)^2 \mathrm{s}^{-1},
\end{align}
when $Q/T<1$ is naively satisfied.

Next we consider $\lambda_{p\nu}$:
\begin{align}
	\label{app:l_pnu}
	\lambda_{p\nu} = \frac{1+3g_A^2}{2\pi^3}G_F^2  Q^5 \hat J(1;\infty),
\end{align}
which is expanded as,
\begin{align}
	\label{l_pnu_int}
	\hat J(1;\infty)&= \int^\infty_1 \sqrt{1-\frac{(m_e/Q)^2}{q^2}}\frac{q^2(q+1)^2\dd q}{(1+\ee^{\frac{Q}{T_\nu}(q+1)})(1+\ee^{-\frac{Q}{T}q})}\nonumber \\
	& \simeq\int^\infty_1 \left( 1-\frac{(m_e/Q)^2}{2q^2}\right) \frac{q^2(q+1)^2\dd q}{1+\ee^{\frac{Q}{T\nu}(q+1)}}.
\end{align}
If the dominant contribution to the integral comes from large $q$, the integral can be rewritten as,
\begin{align}
	&\int^\infty_1 \left( 1-\frac{(m_e/Q)^2}{2q^2}\right) \frac{q^2(q+1)^2\dd q}{1+\ee^{\frac{Q}{T\nu}(q+1)}}\nonumber \\
	&\simeq
	\label{app:l_pnu_int}
	\ee^{- \frac{Q}{T_\nu}} \int^\infty_1 \left( 1-\frac{(m_e/Q)^2}{2q^2}\right) \frac{q^2(q+1)^2\dd q}{1+\ee^{\frac{Q}{T\nu}q}},
\end{align}
where we have used the fact that $1/(1+\ee^{\frac{Q}{T_\nu}(q+1)})\simeq \ee^{-\frac{Q}{T_\nu}}/(1+\ee^{\frac{Q}{T_\nu}q})$ when $q\gg 1$.
Comparing Eqs.~(\ref{app:l_nnu_int}) and (\ref{app:l_pnu_int}), we obtain the approximation form as,
\begin{align}
	\lambda_{p\nu} \simeq 1.63\,\ee^{- \frac{Q}{T_\nu}} \left( \frac{T_\nu}{Q} \right)^3 \left( \frac{T_\nu}{Q} + 0.25 \right)^2 \mathrm{s}^{-1}.
\end{align}

Finally we derive the approximation for $\lambda_{pe}$:
\begin{align}
	\lambda_{pe} = \frac{1+3g_A^2}{2\pi^3}G_F^2  Q^5 \hat J(-\infty;-\frac{m_e}{Q}), 
\end{align} 
which is expanded as, 
\begin{align}
	\label{l_pe_int}
 	\hat J(-\infty;-\frac{m_e}{Q})&= \int^{-\frac{m_e}{Q}}_{-\infty} \sqrt{1-\frac{(m_e/Q)^2}{q^2}} \nonumber \\
	&\times\frac{q^2(q+1)^2\dd q}{(1+\ee^{\frac{Q}{T_\nu}(q+1)})(1+\ee^{-\frac{Q}{T}q})}\nonumber \\
	& \simeq\int^{-\frac{m_e}{Q}}_{-\infty} \left( 1-\frac{(m_e/Q)^2}{2q^2}\right) \frac{q^2(q+1)^2\dd q}{1+\ee^{-\frac{Q}{T}q}} \nonumber \\
	& =\int^\infty_{\frac{m_e}{Q}}\left( 1-\frac{(m_e/Q)^2}{2q^2}\right)\frac{q^2(q-1)^2\dd q}{1+\ee^{\frac{Q}{T}q}}.
\end{align}
Assuming the dominant contribution comes from large $q$, the integral can be written as,
\begin{align}
	&\int^\infty_{\frac{m_e}{Q}}\left( 1-\frac{(m_e/Q)^2}{2q^2}\right) \frac{q^2(q-1)^2\dd q}{1+\ee^{\frac{Q}{T}q}}\nonumber \\
	& \simeq 
	\label{app:l_pe_int}
	\ee^{-\frac{Q}{T}}\int^\infty_{\frac{m_e}{Q}}\left( 1-\frac{(m_e/Q)^2}{2q^2}\right) \frac{q^2(q-1)^2\dd q}{1+\ee^{\frac{Q}{T}(q-1)}},
\end{align} 
where we have used the fact that $1/(1+\ee^{\frac{Q}{T}q})\simeq \ee^{-\frac{Q}{T}}/(1+\ee^{\frac{Q}{T}(q-1)})$ when $q\gg 1$. 
Then, comparing Eqs.~(\ref{app:l_nnu_int_i}) and (\ref{app:l_pe_int}), we can derive,
\begin{align}
	\lambda_{pe}\simeq1.63 \, \ee^{- \frac{Q}{T}} \left( \frac{T}{Q} \right)^3 \left( \frac{T}{Q} + 0.25 \right)^2 \mathrm{s}^{-1}.
\end{align}

\begin{figure}
	\centering
	\includegraphics[width=0.9\hsize]{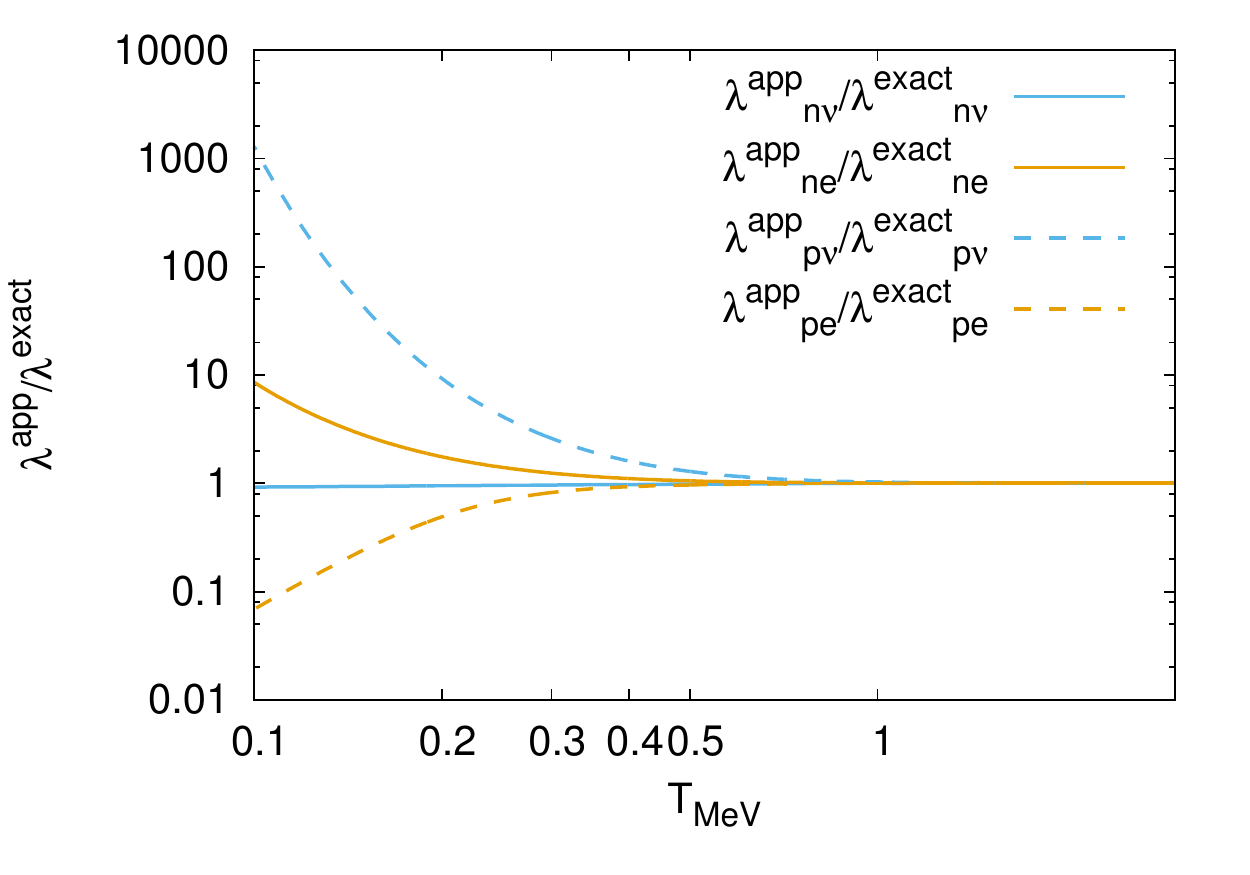}
	\caption{The ratios of the approximated reaction rates $\lambda^{\mathrm{app}}$ to the exact ones $\lambda^{\mathrm{exact}}$. }
	\label{fig:l_compare}
\end{figure}

These approximation forms are valid when $Q/T<1$ is satisfied. 
However, in fact, we can numerically find that these approximation forms are valid even at $T\sim0.4\,\Mev$.
In Fig.~\ref{fig:l_compare}, we plot the ratio of the approximatied reaction rates $\lambda^{\mathrm{app}}$ to the exact ones $\lambda^{\mathrm{exact}}$. 
From this plot, it can be found that the approximation is valid at $T\gtrsim 0.4\,\Mev$.

\begin{widetext}
\section{The detail derivation of the approximated reaction rates with perturbations}
\label{app:taylor_ex}

In this section, we derive the approximated reaction rates $\lambda$'s under the existence of perturbations, Eqs.~(\ref{l_nnu_app})--(\ref{l_pnu_app}).
The approximated rates~(\ref{l_nnu_app_wop})--(\ref{l_pe_app_wop}) are modified by perturbations as,
\begin{align}
	\lambda_{n\nu}(\Theta)&\simeq\int^1_{-1} \frac{\dd\mu}{2}1.63\left( \frac{T_\nu(1+\Theta)}{Q} \right)^3 \left(\frac{T_\nu(1+\Theta)}{Q} + \frac{1}{4} \right)^2 \mathrm{s}^{-1},\\
	\lambda_{ne}(\Theta)&\simeq\int^1_{-1} \frac{\dd\mu}{2}1.63\left( \frac{T(1+\Theta)}{Q} \right)^3 \left(\frac{T(1+\Theta)}{Q} + \frac{1}{4} \right)^2 \mathrm{s}^{-1},\\
	\lambda_{p\nu}(\Theta)&\simeq\int^1_{-1} \frac{\dd\mu}{2}1.63\,\ee^{-\frac{Q}{T_\nu(1+\Theta)}} \left( \frac{T_\nu(1+\Theta)}{Q} \right)^3 \left(\frac{T_\nu(1+\Theta)}{Q} + \frac{1}{4} \right)^2 \mathrm{s}^{-1},\\
	\lambda_{pe}(\Theta)&\simeq\int^1_{-1} \frac{\dd\mu}{2}1.63\,\ee^{-\frac{Q}{T(1+\Theta)}} \left( \frac{T(1+\Theta)}{Q} \right)^3 \left(\frac{T(1+\Theta)}{Q} + \frac{1}{4} \right)^2 \mathrm{s}^{-1}.
\end{align}
Assuming the fluctuation is small, we expand the approximation forms with respect to $\Theta$,  which leads to the approximated reaction rates given by,
\begin{align}
	\lambda_{ne}(\Theta) &\simeq \lambda_{n\nu}(\Theta)\nonumber \\
	&\simeq \int^1_{-1} \frac{\dd\mu}{2} 1.63\left( \frac{T(1+\Theta)}{Q} \right)^3 \left(\frac{T(1+\Theta)}{Q} + \frac{1}{4} \right)^2 \mathrm{s}^{-1}\nonumber \\
	&\simeq\int^1_{-1} \frac{\dd\mu}{2} 1.63 \left( \frac{T}{Q} \right)^3(1+3\Theta +3\Theta^2)\left( \left(\frac{T}{Q}\right)^2 (\Theta^2 + 2\Theta + 1)+\frac{T}{2Q}(1+\Theta) + \frac{1}{16}\right)\mathrm{s}^{-1}\nonumber \\
	&=\int^1_{-1} \frac{\dd\mu}{2} 1.63 \left( \frac{T}{Q} \right)^3(1+3\Theta +3\Theta^2) \left( a+b\Theta + c\Theta^2 \right)\mathrm{s}^{-1}\nonumber \\
	&=\int^1_{-1} \frac{\dd\mu}{2} 1.63\left( \frac{T}{Q} \right)^3(a+(3a+b)\Theta + (3a+3b+c) \Theta^2)\, \mathrm{s}^{-1}\nonumber \\
	&=1.63\left( \frac{T}{Q} \right)^3(a+(3a+b)\Theta_0 + (3a+3b+c) \bar{\Theta}^2)\, \mathrm{s}^{-1}, \\[0.6em]
	\lambda_{pe}(\Theta) &\simeq \lambda_{p\nu}(\Theta) \nonumber \\
	&\simeq \int^1_{-1} \frac{\dd\mu}{2} 1.63\,\ee^{-\frac{Q}{T(1+\Theta)}} \left( \frac{T(1+\Theta)}{Q} \right)^3 \left(\frac{T(1+\Theta)}{Q} + \frac{1}{4} \right)^2  \mathrm{s}^{-1}\nonumber \\
	&\simeq\int^1_{-1} \frac{\dd\mu}{2} 1.63\, \ee^{-\frac{Q}{T}} \left(1+\frac{Q}{T}\Theta +\left[ \frac{1}{2}\left(\frac{Q}{T}\right)^2-\frac{Q}{T}\right]\Theta^2 \right)\nonumber \\
	&\times \left( \frac{T}{Q} \right)^3 (1+3\Theta +3\Theta^2) \left( \left(\frac{T}{Q}\right)^2 (\Theta^2 + 2\Theta + 1)+\frac{T}{2Q}(1+\Theta) + \frac{1}{16}\right)\mathrm{s}^{-1}\nonumber\\
	&=\int^1_{-1} \frac{\dd\mu}{2} 1.63\,\ee^{-\frac{Q}{T}} \left(1+\frac{Q}{T}\Theta +\left[ \frac{1}{2}\left(\frac{Q}{T}\right)^2-\frac{Q}{T}\right] \Theta^2 \right) \left( \frac{T}{Q} \right)^3  
	(1+3\Theta +3\Theta^2) \left( a+b \Theta + c \Theta^2 \right)\mathrm{s}^{-1} \nonumber \\
	&=
	\label{l_pn_app}
	\int^1_{-1} \frac{\dd\mu}{2} 1.63\, \ee^{-\frac{Q}{T}} \left( \frac{T}{Q} \right)^3 \left(a + \left(3a+b+a\frac{Q}{T}\right)\Theta 
	+ \left(a\left[\frac{1}{2}\left(\frac{Q}{T}\right)^2-\frac{Q}{T}\right]+3a+3b+c+\frac{Q}{T}(3a+b)\right)\Theta^2\right)\mathrm{s}^{-1}\nonumber \\	
	&=1.63\, \ee^{-\frac{Q}{T}} \left( \frac{T}{Q} \right)^3 \left(a + \left(3a+b+a\frac{Q}{T}\right)\Theta_0 + \left(a\left[\frac{1}{2}\left(\frac{Q}{T}\right)^2-\frac{Q}{T}\right]+3a+3b+c+\frac{Q}{T}(3a+b)\right)\bar{\Theta}^2\right)\mathrm{s}^{-1},
\end{align}
where we have assumed $T=T_\nu$ and $a$, $b$, and $c$ are defined as,
\begin{align}
	a&= \left(\frac{T}{Q} \right)^2 + \frac{T}{2Q} + \frac{1}{16},\\
	b&= 2\left( \frac{T}{Q} \right)^2 + \frac{T}{2Q}, \\
	c&= \left( \frac{T}{Q} \right)^2.
\end{align}
\end{widetext}


\end{document}